\documentclass [prd,nofootinbib]  {revtex4}
\def\bbbone{{\mathchoice {\rm 1\mskip-4mu l} {\rm 1\mskip-4mu l}
{\rm 1\mskip-4.5mu l} {\rm 1\mskip-5mu l}}}
\begin{document}

\title{%
De Sitter space as an arena for Doubly Special Relativity }
\author{ Jerzy \surname{Kowalski--Glikman}}
\email{jurekk@ift.uni.wroc.pl}
\thanks{Research  partially supported
by the    KBN grant 5PO3B05620.}
\affiliation{ Institute for Theoretical
Physics\\ University of Wroc\l{}aw\\ Pl.\ Maxa Borna 9\\
Pl--50-204 Wroc\l{}aw, Poland}

\begin{abstract}
We show that Doubly Special Relativity (DSR) can be viewed as a theory with energy-momentum space being the four dimensional de Sitter space. Different formulations (bases) of the DSR theory considered so far can be therefore understood as different coordinate systems on this space. The emerging geometrical picture makes it possible to understand the universality of the non-commutative structure of space-time of Doubly Special Relativity. Moreover, it suggests how to construct the most natural DSR bases and makes it possible to address the long standing problem of  total momentum of many particle systems from a different perspective.

\end{abstract}

\maketitle
\section{The DSR theory}

Doubly Special Relativity theory is a new attempt to approach the problem of quantum gravity. This theory was proposed about a year ago by Amelino-Camelia \cite{gac} and is based on two fundamental assumptions: the principle of relativity and the postulate of existence of two observer-independent  scales, of speed identified with the speed of light $c$\footnote{In what follows we set $c=1$.}, and of mass $\kappa$ (or length $\ell=1/\kappa$) identified with the Planck mass. There are several theoretical indications that such a theory may replace Special Relativity as a theory of relativistic kinematics of probes whose energies  are close to the Planck scale. First of all both loop quantum gravity and string theory indicate appearance of the minimal length scale. It is therefore not impossible that this scale would be  present in description of ultra high energy kinematics even in the regime, in which gravitational effects are negligible. Secondly, in both inflationary cosmology \cite{cosmo} and in black hole physics \cite{bh} one faces the conceptual ``trans-Planckian puzzle'' of ordinary physical quanta being blue shifted up to the Planck energies, which as advocated by many can be solved by assuming deviation from the standard dispersion relation at high energies, and thus deviation from the standard relativistic kinematics. It should be also stressed that some Doubly Special Relativity models might provide a resolution of observed anomalies in astrophysical data \cite{astro}. Moreover, predictions of the DSR scenario might be testable in forthcoming quantum gravity experiments \cite{qg}.

Soon after appearance of the papers \cite{gac} it was realized \cite{jkgminl}, \cite{rbgacjkg} that the so called $\kappa$-Poincar\'e algebra in the bicrossproduct basis \cite{kappaP} provides an example of the energy-momentum sector of DSR theory\footnote{More recently another form of the DSR theory was presented by \cite{JoaoLee}. Relations between different forms of DSR were discussed in  \cite{juse}, \cite{lunoDSR}.}. This algebra consists of undeformed Lorentz generators
$$
[M_i, M_j] = i\, \epsilon_{ijk} M_k, \quad [M_i, N_j] = i\, \epsilon_{ijk} N_k,
$$
\begin{equation}\label{1}
  [N_i, N_j] = -i\, \epsilon_{ijk} M_k,
\end{equation}
the standard action of rotations on momenta
\begin{equation}\label{2}
  [M_i, p_j] = i\, \epsilon_{ijk} p_k, \quad [M_i, p_0] =0,
\end{equation}
along with the deformed action of boosts on momenta
\begin{equation}\label{3}
   \left[N_{i}, p_{j}\right] = i\,  \delta_{ij}
 \left( {\kappa\over 2} \left(
 1 -e^{-2{p_{0}/ \kappa}}
\right) + {1\over 2\kappa} \vec{p}\,{}^{ 2}\, \right) - i\,
{1\over \kappa} p_{i}p_{j} 
\end{equation}
governed by the observer-independent mass scale $\kappa$.

The algebra (\ref{1}--\ref{3}) is, of course, not unique. The presence of the second observer-independent scale $\kappa$ makes it possible to consider transformations to another DSR basis, in which (\ref{1}) holds, an thus the Lorentz subalgebra is left unchanged, but one introduces new momentum variables
\begin{equation}\label{4}
 p'_0 = f(p_0, \vec{p}\,{}^{ 2}; \kappa), \quad p'_i = g(p_0, \vec{p}\,{}^{ 2}; \kappa)p_i.
\end{equation}
By construction $p'_0$ and $p'_i$ transform under rotations as scalar and vector, respectively. The functions $f$ and $g$ are assumed to be analytical in the variables $p_0$ and $\vec{p}\,{}^{ 2}$, and in order to guarantee the correct low energy behavior one assumes that for $\kappa\rightarrow\infty$
\begin{equation}\label{4a}
f(p_0, \vec{p}\,{}^{ 2}) \approx p_0 + O(1/\kappa), \quad g(p_0, \vec{p}\,{}^{ 2}) \approx 1 + O(1/\kappa).
\end{equation}
It can be shown \cite{jkgsn} that also vice-versa, any deformed Poincar\'e algebra with undeformed Lorentz sector and standard action of rotations, which has the standard Poincar\'e algebra as its $\kappa\rightarrow\infty$ limit can be related to the algebra (\ref{1}--\ref{3}) by transformation of the form (\ref{4}). One should note in passing that this means in particular that any modified dispersion relation considered in the context ``trans Planckian problem'' can be extended to a DSR theory, and thus does not need to lead to breaking of Lorentz symmetry.

The algebra (\ref{1}--\ref{3}) does not furnish the whole physical picture of DSR theory. To describe physical processes we need also a space-time sector of this theory. The question arises as to if it is possible to construct this sector from the energy-momentum sector. The answer turns out to be affirmative if one extends the energy-momentum DSR algebra to the quantum (Hopf) algebra. It was shown in \cite{jkgsn} that such an extension is possible in the case of any DSR algebra, in particular, for the algebra (\ref{1}--\ref{3}) one gets the following expressions for the co-product
\begin{equation}\label{5}
 \Delta(p_{i}) = p_{i}\otimes \bbbone +
e^{-{p_{0}/ \kappa}} \otimes p_{i}\, ,
\end{equation}
\begin{equation}\label{6} 
\Delta(p_{0}) = p_{0}\otimes \bbbone +  \bbbone \otimes p_{0}\, ,
\end{equation}
\begin{equation}\label{7}
 \Delta(N_{i}) = N_{i}\otimes \bbbone  +
e^{-{p_{0}/ \kappa}}\otimes N_{i} + {1\over \kappa}
\epsilon_{ijk}p_{j}\otimes M_{k}
\end{equation}
(the co-product for rotations is trivial.) Then one makes use of the unique ``Heisenberg double'' prescription \cite{crossalg} in order to get the following commutators 
\begin{equation}\label{8}
[p_0, x_0] = i, \quad [p_i, x_j] = -i \, \delta_{ij},
\quad
 [p_i, x_0] = -\frac{i}\kappa\, p_i.
\end{equation}
By using the same method one finds also that the space-time of DSR theory is non-commuting
\begin{equation}\label{9}
[x_0, x_i] = -\frac{i}\kappa\, x_i.
\end{equation}
and that position operators transform under boosts in the following way \cite{jkgsn}, \cite{crossalg}
\begin{equation}\label{11}
[N_i, x_j] = i \delta_{ij} x_0 - \frac{i}\kappa\, \epsilon_{ijk} M_k, \quad [N_i, x_0] = i x_{i} - \frac{i}\kappa\, N_i.
\end{equation}
($x_0$ and $x_i$ transform as scalar and vector under rotations.) 

It was proved in \cite{jkgsn} that both the space-time non-commutativity (\ref{9}) and the form of the boost action on position operators (\ref{11}) is universal for all DSR theories, i.e., it is invariant of energy-momentum transformations (\ref{4}), (\ref{4a})\footnote{Strictly speaking this result follows if one uses the Heisenberg double construction. However it should be stressed that this is the only known way do derive the space-time sector of DSR from the energy-momentum one, which in turn means that this is the only known  construction of the complete DSR theory.}. This raises a suspicion that all the DSR theories might be in fact equivalent physically equivalent and different bases in momentum space play a role similar to those played by different coordinate systems on given space-time of General Relativity. In this paper we will not present any arguments supporting this claim (we will discuss this in another paper). Instead we will try to clarify the emerging picture of DSR theories as a whole by trying to look at them from a different perspective. All the past developments in DSR took place in the framework of the algebraic approach (commutational algebras, Casimir operators, etc.) It turns out that the DSR theories posses an appealing geometric picture as well, and we devote this paper to investigations of geometry of the DSR theory.

\section{DSR algebra and de Sitter space}

Among infinitely many DSR bases, related to each other by transformation (\ref{4}), (\ref{4a}) the most important for our purposes is the basis proposed long ago by Snyder \cite{snyder}. In this basis the action of Lorentz algebra on  energy-momentum  sector is classical, i.e., 
\begin{equation}\label{12}
  [N_i, P_j] = i \delta_{ij} P_0 , \quad [N_i, P_0] = i P_{i},
\end{equation}
while for positions we have the universal algebra (\ref{11}). Moreover
$$ [P_i, {X}_j] = - i \delta_{ij}\left(1 + \frac1{\kappa}\, P_0\right) -  \frac{i}{\kappa^2}\, P_i P_j, $$
$$ [{X}_i, P_0] =  \frac{i}{\kappa}\, P_i + \frac{i}{\kappa^2}\, P_i P_0, $$
$$ [{X}_i, P_0] =  \frac{i}{\kappa}\, P_i +\frac{i}{\kappa^2}\, P_i P_0,$$
\begin{equation}\label{13}
 [P_0, {X}_0] = i\left(1 - \frac{1}{\kappa^2}\, P_0^2\right),
\end{equation}
and again the commutator of space and time is given by (\ref{9}).
\newline

As it stands, the algebra (\ref{9}), (\ref{11}), (\ref{12}), (\ref{13}) looks like a particular DSR basis. The important observation, following the original construction of Snyder is that the momenta $P_0$ and $P_i$ can be viewed as coordinates on de Sitter space.  Indeed, let de Sitter space be defined by equation
\begin{equation}\label{17}
 -\eta_0^2 + \eta_1^2+ \eta_2^2+ \eta_3^2+ \eta_4^2 =\kappa^2,
\end{equation}
and let us define the coordinates
\begin{equation}\label{18}
 P_\mu = \kappa\frac{\eta_\mu}{\eta_4}, \quad \mu = (0, \ldots, 3).
\end{equation}
It is clear that the coordinates $P_\mu$  cover only half of the whole de Sitter space (the points $(\eta_\mu, \eta_4)$ and $(-\eta_\mu, -\eta_4)$ are identified in these coordinates.) If one now derives the form of generators of $SO(4,1)$ symmetry of de Sitter space in these coordinates, such that $M_i$, $N_i$ belong to its $SO(3,1)$ subalgebra, while $X_\mu$ are the remaining four generators belonging to the quotient of two algebras $SO(4,1)/SO(3,1)$, one finds that they satisfy the $SO(4,1)$ relations (\ref{1}), (\ref{9}), (\ref{11}) as well as the cross relations (\ref{13}).

This simple observation clarifies the universal status of the algebra satisfied by positions and boost and rotation generators in any DSR basis. To understand this let us look at the DSR theory from geometric perspective suggested by Snyder's construction. From this viewpoint the space of momenta is not a flat space, as in Special Relativity, but a curved, maximally symmetric space of constant curvature. The fact that we need a maximally symmetric space is related, of course to the fact that only such space has the required number of symmetry generators, namely six ``rotations'' identified with Lorentz transformations and four ``translations'' in the energy-momentum space, which can be identified with (non-commutative) positions. It is well known that there are only three families of maximally symmetric spaces: de Sitter, Anti de Sitter and the flat space of constant positive, negative, and zero curvature, respectively. Next, it is clear that
 even though  on the (momentum) de Sitter space one can introduce arbitrary coordinates (each corresponding to a particular DSR basis), the form of symmetries of this space does not, of course depend on the form of the coordinate system (recall that de Sitter space is the quotient space $SO(4,1)/SO(3,1)$). In other words the momentum sectors of various DSR theories are in one to one correspondence with differential structures that can be built on de Sitter space, while the structure of the positions/boosts/rotations, being related to the symmetries of this space is, clearly, diffeomorphism-invariant. One should note also that the fact that Heisenberg double construction leads to algebraic structure consistent with the geometric picture of the DSR theory indicates that this is the right way of construction of the space-time sector of this theory.

At this point a question arises, namely if the coordinates (\ref{18}) are the most natural ones from the geometric perspective. Indeed in these coordinates the cross relations (\ref{13})  the physical meaning of positions as generators of translations in energy-momentum space is far from being manifest. It is therefore useful to try to construct a coordinate system in which the physical role played by positions exhibits itself in a more clear way. This can be done as follows. Consider the point ${\cal O}$ in de Sitter momentum space with coordinates $(\eta_\mu, \eta_4) = (0,\kappa)$. This point corresponds to the zero momentum in the coordinate system (\ref{18}) and we assume that it corresponds to zero momentum state in any coordinates as well. Geometrically this assumption corresponds to  defining the preferred point in de Sitter space, but, of course it is well motivated physically. Since de Sitter space equals $SO(4,1)/SO(3,1)$, the stability group of this point is just $SO(3,1)$ of $M_i$ and $N_i$, and the remaining four generators of $SO(4,1)$, ${ X}_\mu$ can be used to define points on de Sitter space as follows. One observes that the group elements  $\exp(i{\cal P}_0 { X}_0)$, $\exp(i{\cal P}_i { X}_i)$ have  natural interpretation of ``large translations'' in the momentum de Sitter space (since this space is curved the translations cannot commute.) Indeed 
$$
\exp(i{\cal P}^{(1)}_0 { X}_0)\, \exp(i{\cal P}^{(2)}_0 { X}_0)=\exp(i({\cal P}^{(1)}_0+{\cal P}^{(2)}_0) { X}_0), \quad
\exp(i{\cal P}^{(1)}_i { X}_i)\, \exp(i{\cal P}^{(2)}_i { X}_i)=\exp(i({\cal P}^{(1)}_i+{\cal P}^{(2)}_i) { X}_i)
$$
while
\begin{equation}\label{19x}
\exp(i{\cal P}_0 { X}_0)\, \exp(i{\cal P}_i { X}_i)\, \exp(-i{\cal P}_0 { X}_0) =\exp\left(ie^{-{\cal P}_0/\kappa}{\cal P}_i { X}_i\right).
\end{equation}
Now one can define the natural coordinates on this space by  labelling the point 
\begin{equation}\label{19y}
\wp \equiv {\cal G}({\cal P}_0,{\cal P}_i)\, {\cal O} = \exp(i{\cal P}_0 { X}_0)\,\exp(i{\cal P}_i { X}_i)\, {\cal O}
\end{equation} 
with coordinates ${\cal P}_\mu$. 

With the help of explicit  form of $X_\mu$ generators
in the matrix representation
\begin{equation}\label{19a}
X_0 = \frac{i}{\kappa} \,\left(\begin{array}{ccc}
  0 & \mathbf{0} & 1 \\
  \mathbf{0} & \mathbf{0} & \mathbf{0} \\
  1 & \mathbf{0} & 0 
\end{array}\right) \quad
\vec{X} = \frac{i}{\kappa} \,\left(\begin{array}{ccc}
  0 & {\vec{\epsilon}\,{}^T} &  0\\
  \vec{\epsilon} & \mathbf{0} & \vec{\epsilon} \\
  0 & -\vec{\epsilon}\,{}^T & 0 
\end{array}\right),
\end{equation}
where $\vec{\epsilon}$ is a  three-vector with one non-vanishing unit component and $\vec{\epsilon}^T$ the associated transposed vector, one finds
\begin{equation}\label{19b}
\exp(i{\cal P}_0 { X}_0)  = \left(\begin{array}{ccc}
  \cosh \frac{{\cal P}_0}\kappa & \mathbf{0} & -\sinh \frac{{\cal P}_0}\kappa \\
  \mathbf{0} & \bbbone & \mathbf{0} \\
 -\sinh \frac{{\cal P}_0}\kappa  & \mathbf{0} &  \cosh \frac{{\cal P}_0}\kappa
\end{array}\right) \quad
\exp(i{\cal P}_i { X}_i) = \left(\begin{array}{ccc}
  1+ \frac{\vec{{\cal P}}\,{}^2}{2\kappa^2} & -\frac{\vec{{\cal P}}^T}{\kappa} &  \frac{\vec{{\cal P}}\,{}^2}{2\kappa^2}\\
  -\frac{\vec{{\cal P}}}{\kappa} & \bbbone & -\frac{\vec{{\cal P}}}{\kappa} \\
  -\frac{\vec{{\cal P}}\,{}^2}{2\kappa^2} & \frac{\vec{{\cal P}}^T}{\kappa} & 1 - \frac{\vec{{\cal P}}\,{}^2}{2\kappa^2} \end{array}\right),
\end{equation}
and the coordinates $({\cal P}_\mu)$ label the point $\wp = (\eta_0, \ldots,\eta_4)$ with 
\begin{eqnarray}
{\eta_0} &=& -\kappa\, \sinh \frac{{\cal P}_0}\kappa + \frac{\vec{{\cal P}}\,{}^2}{2\kappa}\, e^{ \frac{{\cal P}_0}\kappa} \nonumber\\
\eta_i &=&  - {\cal P}_i  \nonumber\\
{\eta_4} &=& \kappa\, \cosh \frac{{\cal P}_0}\kappa  -\frac{\vec{{\cal P}}\,{}^2}{2\kappa} \, e^{ \frac{{\cal P}_0}\kappa}   \label{20}
\end{eqnarray}
From these expressions and from the relations
$$
[X_0,\eta_4] = \frac{i}\kappa\, \eta_0, \quad [X_0,\eta_0] = \frac{i}\kappa\, \eta_4, \quad [X_0,\eta_i] = 0,
$$
$$
[X_i, \eta_4] = [X_i, \eta_0] =\frac{i}\kappa\, \eta_i, \quad [X_i, \eta_j] = \frac{i}\kappa\, \delta_{ij}(\eta_0 - \eta_4),
$$
$$
[N_i,\eta_0] = i \eta_i, \quad [N_i,\eta_j] = i \, \delta_{ij}\, \eta_0, \quad [N_i, \eta_4] = 0
$$
one can read off the form of non-vanishing cross commutators
\begin{equation}\label{21}
[{\cal P}_0, X_0] = i, \quad [{\cal P}_i, X_j] = - i\, \delta_{ij} \left(1 - \frac{\vec{{\cal P}}\,{}^2}{\kappa^2}\right)\, e^{{\cal P}_0/\kappa}, \quad [{\cal P}_0, X_i] = -\frac{2i}\kappa\, {\cal P}_i\, e^{{\cal P}_0/\kappa}
\end{equation}
as well as the action of boosts on momenta
\begin{equation}\label{22}
[N_i, {\cal P}_j] =  i\, \delta_{ij}\left(\kappa\sinh \frac{{\cal P}_0}\kappa - \frac{\vec{{\cal P}}\,{}^2}{2\kappa}\, e^{{\cal P}_0/\kappa}\right)
,\quad [N_i, {\cal P}_0] = i {\cal P}_i\, e^{{\cal P}_0/\kappa}.
\end{equation}
The quadratic Casimir of the algebra (\ref{22}) has the form
\begin{equation}\label{23}
 {\cal C} = -\left(2\kappa\sinh\frac{{\cal P}_0}{2\kappa}\right)^2 + \vec{{\cal P}}\,{}^2\, e^{{\cal P}_0/\kappa}
\end{equation}
and, remarkably, is of the same form as the quadratic Casimir for the bicrossproduct basis (\ref{1}--\ref{3}).

Alternatively, one can choose the prescription $$\tilde\wp \equiv \tilde{\cal G}({\cal P}_0,{\cal P}_i)\, {\cal O} = \exp(i{\cal P}_i { X}_i)\,\exp(i{\cal P}_0 { X}_0)\, {\cal O},$$ in which case one obtains (cf.~(\ref{19x}))
\begin{eqnarray}
{\eta_0} &=& -\kappa\, \sinh \frac{{\cal P}_0}\kappa + \frac{\vec{{\cal P}}\,{}^2}{2\kappa}\, e^{ -\frac{{\cal P}_0}\kappa} \nonumber\\
\eta_i &=&  - {\cal P}_i \, e^{ -\frac{{\cal P}_0}\kappa} \nonumber\\
{\eta_4} &=& \kappa\, \cosh \frac{{\cal P}_0}\kappa  -\frac{\vec{{\cal P}}\,{}^2}{2\kappa} \, e^{ -\frac{{\cal P}_0}\kappa}   \label{24}
\end{eqnarray}
in which case the  cross commutators read
\begin{equation}\label{25}
[{\cal P}_0, X_0] = i, \quad [{\cal P}_i, X_0] = \frac{i}\kappa\, {\cal P}_i, \quad [{\cal P}_i, X_j] = - i\, \delta_{ij} \, e^{2{\cal P}_0/\kappa} +\frac{i}{\kappa^2}\left(\vec{{\cal P}}\,{}^{ 2}\, \delta_{ij} - 2 {\cal P}_{i}{\cal P}_{j}\right), \quad [{\cal P}_0, X_i] = -\frac{2i}\kappa\, {\cal P}_i
\end{equation}
Remarkably, the action of boosts on momenta has the form
\begin{equation}\label{26}
[N_i, {\cal P}_j] =  i\,  \delta_{ij}
 \left( {\kappa\over 2} \left(e^{2{{\cal P}_{0}/ \kappa}}
- 1 
\right) - {1\over 2\kappa} \vec{{\cal P}}\,{}^{ 2}\, \right) + i\,
{1\over \kappa} {\cal P}_{i}{\cal P}_{j} 
,\quad [N_i, {\cal P}_0] = i {\cal P}_i.
\end{equation}
Which is nothing but the boost action in the bicrossproduct basis with $\kappa$ replaced with $-\kappa$. In this case the quadratic Casimir  has the form
\begin{equation}\label{27}
 {\cal C} = -\left(2\kappa\sinh\frac{{\cal P}_0}{2\kappa}\right)^2 + \vec{{\cal P}}\,{}^2\, e^{-{\cal P}_0/\kappa}
\end{equation}

\section{Geometric rule for total momentum}

One of the major unsolved problem of the DSR theory is how to construct a physical quantity having an interpretation of total momentum of many particle system. Currently there are two proposals seriously considered  in the literature, each of whose is hardly acceptable. The first advocated in \cite{lunoDSR} makes use of the co-product structure of the DSR theory. This proposal has two unphysical features, namely that the total momentum is not a symmetric function of momenta of particles composing the system, and second that it preserves the mass Casimir, which means that the square of three-momentum of any composite system cannot be larger that $\kappa^2c^2$ which obviously contradicts everyday experiment (since $\kappa  c \sim 10^{-28}\, kg\, m/s$.) The second proposal has been recently put forward in \cite{jv}\footnote{This proposal has been first spelled out in \cite{lunoDSR} in the language of symmetric co-product for non-lineal realization of classical Poincar\'e algebra. A modified variant of this idea has been recently presented in \cite{jl2}.}. The idea is to  transform particles momenta from a particular DSR basis to the classical one (i.e., the basis in which the Lorentz transformations of momenta are classical, as in (\ref{12})), to make use of the standard Special Relativity summation rule to get the total momentum, and inverse transform it to the DSR basis of interest. The problem with this approach is at least twofold. First, there is the whole class of transformation from given a DSR basis to the classical one. Second, it seems far from being obvious that in the DSR theory the rule of composition of momenta is to be identical to the one of Special Relativity even in the classical basis. The point is that in Special Relativity there is no alternative to the standard addition rule because there is no mass scale to play with. If the scale is present there are, of course, infinitely many consistent rules of  addition of momenta.

It is interesting to note that the group-theoretical picture revealed in the analysis presented in the preceding section leads in a natural way to the ``co-product'' summation rule. To see this let us take the group summation rule (cf.~(\ref{19y}))
\begin{equation}\label{28}
 \mathcal{G}\left({\cal P}_0^{(1)} \oplus {\cal P}_0^{(2)},{\cal P}_i^{(1)} \oplus {\cal P}_i^{(2)}\right) =\mathcal{G}\left({\cal P}_0^{(1)} ,{\cal P}_i^{(1)} \right)\,\mathcal{G}\left({\cal P}_0^{(2)} ,{\cal P}_i^{(2)} \right). 
\end{equation}
Using (\ref{19x}) one easily finds that
\begin{equation}\label{29}
 {\cal P}_0^{(1)} \oplus {\cal P}_0^{(2)} = {\cal P}_0^{(1)} + {\cal P}_0^{(2)}, \quad {\cal P}_i^{(1)} \oplus {\cal P}_i^{(2)} = e^{-{\cal P}_0^{(2)}/\kappa}\, {\cal P}_i^{(1)} + {\cal P}_i^{(2)}
\end{equation}
which is indeed the ``co-product'' sum of momenta \cite{lunoDSR} (cf.~(\ref{7})). This observation makes it possible to understand the lack of symmetry of this summation prescription as a result of non-abelian structure of the group of translation acting on de Sitter space of momenta.

The geometric picture of the DSR theory makes it possible to introduce another rules for momenta addition. To understand it, consider the standard special relativistic case. To add two momenta $p^{(1)}$ and $p^{(2)}$ one first constructs the  line segments ${\cal L}_{1/2}$ in the momentum space connecting the point corresponding to zero momentum ${\cal O}$ with the points $p^{(1)}$ and $p^{(2)}$, respectively. Next one parallel transports the tangent vectors to the segment ${\cal L}_{1/2}$ to the points  $p^{(2/1)}$ and constructs the straight lines, to whose the resulting vectors is tangent. The intersection of these lines provides the point in the momentum space, whose coordinates correspond to the total momentum. The parallelogram method described above can be readily extended to the case of momentum manifold being curved. This amounts only in replacing straight line segments by geodesic segments, and making use of the standard definition of parallel transport of vectors along geodesics in Riemannian geometry. This definition of total momentum provides the total momentum that is manifestly symmetric function of the momenta of (two) constituents. It is an open question if this construction can be extended to the system of more than two particles in an associative way.

\section{Conclusions}

The main result of this paper is that any DSR theory can be regarded as a particular coordinate system on de Sitter space of momenta. In addition  the Lorentz transformations have interpretations of stabilizers of the zero-momentum point in the de Sitter space, while positions are identified with the remaining four generators of $SO(4,1)$. This observation leads naturally to the claim that one should formulate physical models on this momentum space in such a way that physical quantities would be independent of the choice of DSR basis, i.e., independent of coordinate system employed. Naturally such  ``general coordinate invariance in the momentum space'' principle is strongly motivated by the equivalence principle and general coordinate invariance of general relativity. It seems that the only reasonable way to check if this principle can be built into physical theory is to try to construct a field theory on curved momentum space in the manifestly coordinate-invariant way. This problem is currently under investigation.

\section*{Acknowledgement} The idea to investigate de Sitter structure of the momentum space of DSR theories arose in the course of many discussions with Giovanni Amelino-Camelia during my visit to  Rome.  I would like to thank him as well as the Department of Physics of the University of Rome ``La Sapienza'' for their worm hospitality during my visit.

\end{document}